\documentclass[prb,superscriptaddress,twocolumn,amsmath,amssymb,10pt,nofootinbib]{revtex4}
\usepackage{graphicx}
\usepackage{dcolumn}
\usepackage{bm}
\usepackage{color}
\begin{document}

\title{Photonic quantum technologies}

\author{Jeremy L. O'Brien}
\affiliation{Centre for Quantum Photonics, H. H. Wills Physics Laboratory \& Department of Electrical and Electronic Engineering, University of Bristol, Merchant Venturers Building, Woodland Road, Bristol, BS8 1UB, UK}

\author{Akira Furusawa}
\affiliation{Department of Applied Physics and Quantum Phase Electronics Center, School of Engineering, The University of Tokyo, 7-3-1 Hongo, Bunkyo-ku, Tokyo 113-8656, Japan
}%

\author{Jelena Vu\v{c}kovi\'{c}}
\affiliation{Department of Electrical Engineering and Ginzton Laboratory, Stanford University, Stanford CA 94305}
\date{\today}
\begin{abstract}%
The first quantum technology, which harnesses uniquely quantum mechanical effects for its core operation, has arrived in the form of commercially available quantum key distribution systems that achieve enhanced security by encoding information in photons such that information gained by an eavesdropper can be detected. Anticipated future quantum technologies include large-scale secure networks, enhanced measurement and lithography, and quantum information processors, promising exponentially greater computation power for particular tasks. Photonics is destined for a central role in such technologies owing to the need for high-speed transmission and the outstanding low-noise properties of photons. These technologies may use single photons or quantum states of bright laser beams, or both, and will undoubtably apply and drive state-of-the-art developments in photonics.
\end{abstract}

\maketitle

The theory of quantum mechanics was developed at the beginning of the twentieth century to better explain the spectra of light emitted by atoms. At the time, many famously believed that all of Physics was close to finalized, with a few remaining anomalies  to be ironed out. The full theory emerged as a completely unexpected description of how the world works at a fundamental level: It painted a picture  that was fundamentally probabilistic, where a single object could be in two places at once---\emph{superposition}---and that two objects in remote locations could be instantaneously connected---\emph{entanglement}.  These unusual properties have been directly observed and quantum mechanics remains  the most successful theory humankind has developed in terms of the precision of its predicitions. Today we are learning how to harness these `bizarre' quantum effects to realize profoundly new `quantum' technologies. 

Quantum information science \cite{nielsen} has emerged over the last decades to address the question `Can we gain new functionality and power by harnessing quantum mechanical effects by storing, processing and transmitting information encoded in inherently quantum mechanical systems?' Fortunately the answer is yes. Quantum information is both a fundamental science and a progenitor of new technologies: already several commercial quantum key distribution systems that offer enhanced security by communicating information encoded in quantum systems are available \cite{gi-nphot-1-165}. 
It is anticipated that such systems will be extended to quantum communication networks, providing security based on the laws of physics. Perhaps the most profound (and distant) anticipated future technology is a quantum computer that promises exponentially faster operation for particular tasks \cite{nielsen}, 
including factoring, database searches, and simulating important quantum systems, which one day may have relevance to technologically important materials. Quantum metrology \cite{gi-sci-306-1330} aims to harness quantum effect in measurement to achieve the highest precision allowed by nature, while quantum lithography aims to use quantum states of light to define features smaller than the wavelength \cite{bo-prl-85-2733}.

There are a number of physical systems being pursued for these future technologies \cite{nielsen}, however, quantum states of light appear destined for a central role: light is a logical choice for quantum communication, metrology and lithography, and is a leading approach to quantum information processing (QIP). Photonic quantum technologies have their origin in the fundamental science of quantum optics, which itself has been a testing ground for the ideas of quantum information science. For example, quantum entanglement was 
experimentally tested using photons generated from atomic cascades in the 1970s and early 1980s \cite{fr-prl-28-938,as-prl-47-460}. 
In the late 1980s and 1990s, the non-linear process of `spontaneous parametric downconversion' (SPDC) 
was shown to be a convenient source of pairs \cite{kw-prl-75-4337} of photons for such fundamental experiments 
and to generate quantum states of a bright laser beam---`squeezed states' \cite{ou-prl-68-3663}. SPDC has been used for many fundamental quantum information tasks, including quantum teleportation \cite{bo-nat-390-575, Furusawa98}. 
Similarly, the interaction of single photons with single atoms in an optical cavity---cavity quantum electrodynamics (QED)---has been a rich field of fundamental science with major applications to photonic quantum technologies \cite{tu-prl-75-4710}

While we don't yet know exactly what form future quantum technologies will take, it seems likely that quantum information will be transmitted in quantum states of light, and that some level of information processing will be implemented on these quantum states. It also seems clear that if we are to realize these technologies we will need to harness and ultimately drive some of the latest developments in the conventional field of photonics.

\begin{figure}
\begin{center}
\includegraphics*[width=8.5cm]{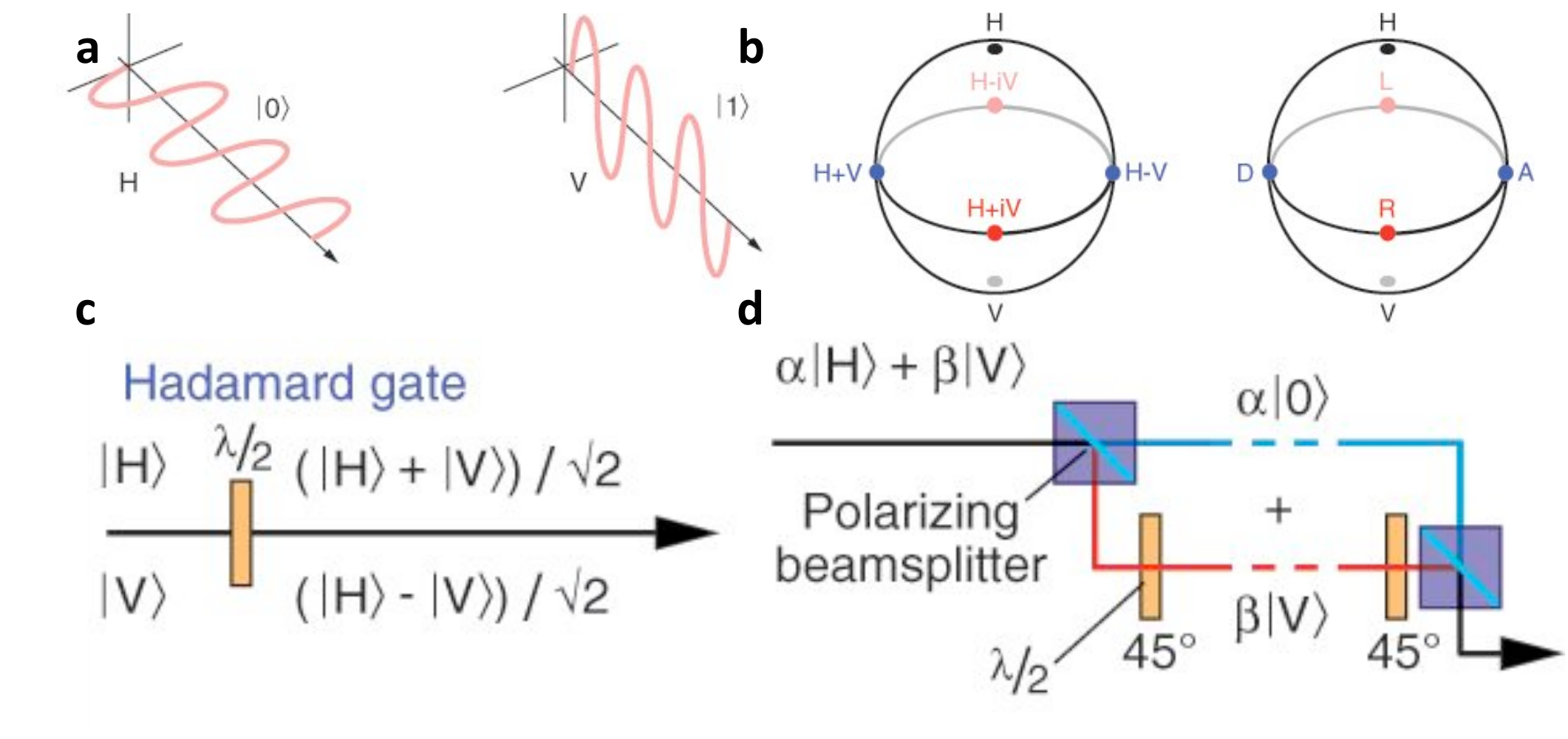}
\vspace{-0.5cm}
\caption{
{\bf a}, A qubit can be encoded as the polarization of a single photon. {\bf b} An arbitrary state of a qubit can be represented on the Poincare\'{e} or Bloch sphere. {\bf c} A half wave plate ($\lambda/2$) can be used to rotate this polarisation. {\bf d} A polarisation encoded qubit can be interconverted to a path enecoded qubit via a polarising beam splitter.
}
\label{schematic}
\end{center}
\vspace{-0.7cm}
\end{figure}

\vspace{12pt}
\noindent\textbf{Secure communication with photons}\\
Light speed transmission and low noise properties make photons indispensable for quantum communication---transferring a quantum state from one place to another \cite{gi-rmp-74-145}. A quantum bit (or qubit) of information can be encoded in any of several degrees of freedom---polarization, path, time-bin \emph{etc}. Manipulation at the single photon level is usually straightforward---using birefringent waveplates in the case of polarization for example (Fig. \ref{schematic}). 

This ability to transfer quantum states between remote locations can be used to greatly enhance communication security. 
We can use the fundamental fact that any measurement of a quantum system necessarily disturbs that system to reliably detect the presence of an eavesdropper. Several commercially available quantum key distribution (QKD) systems operate on this principle (comprehensive reviews on the topic can be found in Refs. \onlinecite{gi-nphot-1-165}, for example). These QKD systems currently rely on attenuated laser pulses rather than single photons---an approach that has been shown to be sufficient for point to point applications \cite{gi-nphot-1-165}. However, attenuation of these weak laser pulses over transmission in fibre, or free space, currently limits the range of such systems to 100's km (quantum states of bright light are described below).

The advanced state of our modern communication systems owes much to the Erbium Doped Fiber Amplifier (EDFA) for it's ability to amplify optical signals as they propagate over long distances of optical fibre. Unfortunately, amplification of a quantum signal is not so straightforward: measurement of the quantum state of the signal destroys the information (the same disturbance that enables detection of an eavesdropper). A major challenge is to realise a quantum repeater that is able to store quantum information and implement entangling measurements. Ultimately, sophisticated quantum networks will likely require nodes with a small-scale version of the quantum information processors described below.

\vspace{12pt}
\noindent\textbf{Quantum information processing}\\
The requirements for realizing a quantum computer are confounding: scalable physical qubits---two state quantum systems---that can be well isolated from the environment, but also initialised, measured, and controllably interacted to implement a universal set of quantum logic gates \cite{di-sm-23-419}. Despite these great challenges, a number of physical implementations are being pursued, including nuclear magnetic resonance, ion, atom, cavity quantum electrodynamics, solid state, and superconducting  systems. Over the last few years single  photons have emerged as a leading approach \cite{ob-sci-318-1567}.

\begin{figure}
\includegraphics*[width=9cm]{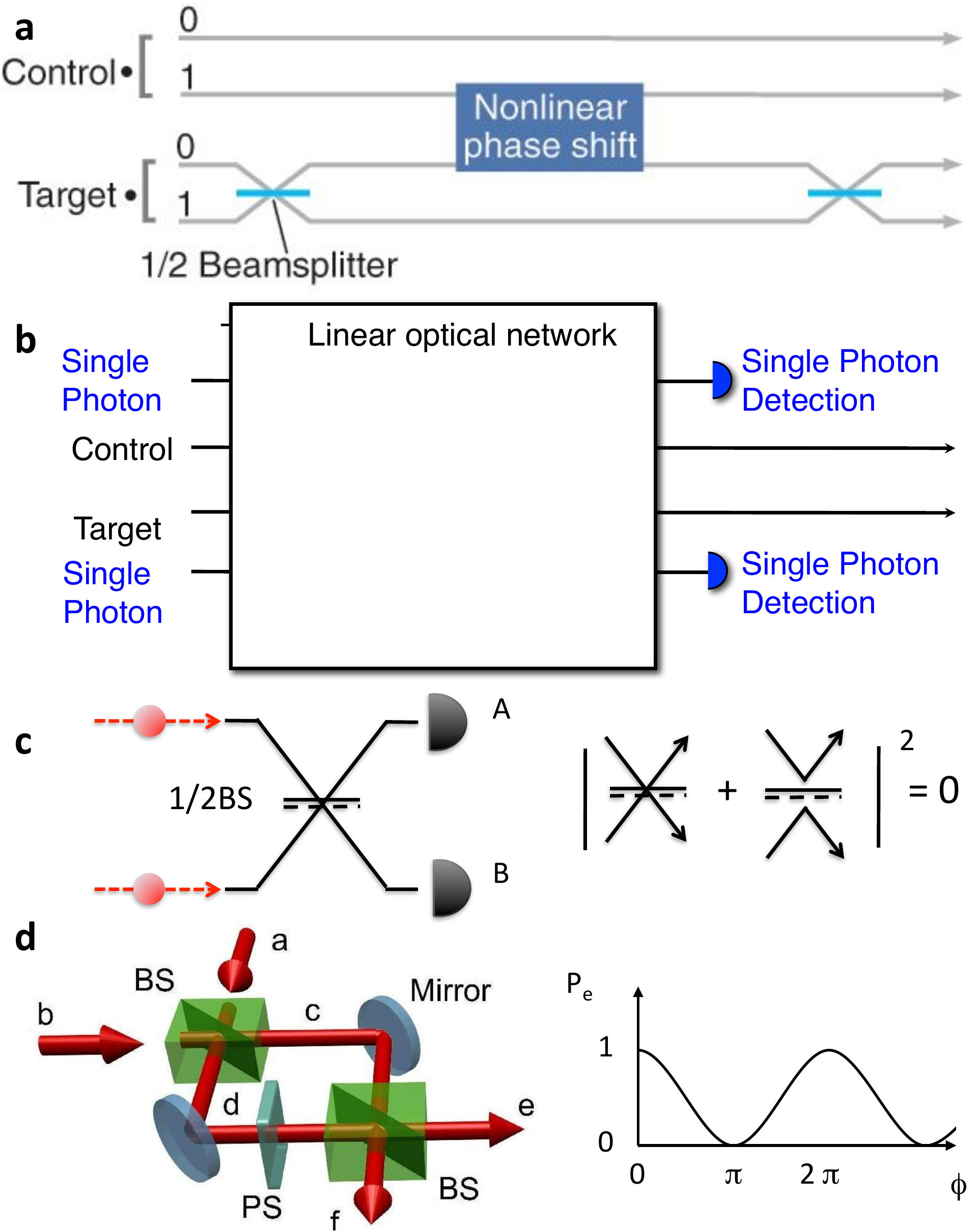}
\caption{An optical controlled-NOT gate. {\bf a}, Schematic of a possible realization of an optical CNOT gate. {\bf b}, Schematic of the KLM scheme. {\bf c}, Quantum interference of photons: two photons arriving simultaneously at a beamsplitter both leave in the same output mode with certainty because quantum interference of the probability amplitudes to detect a photon at A and at B destructively interfere. {\bf d}, A Mach-Zehnder interferometer. The sensitivity with which the phase $\phi$ can be measured is related to the gradient of the interference fringe.}
\vspace{-0.5cm}
\label{cnot}
\end{figure}

A major difficulty for optical quantum information processing (QIP) is in realizing two-qubit entangling logic gates. The canonical example is the controlled-NOT gate (CNOT), which flips the state of a target qubit conditional on a control qubit being in the logical state ``1"---the quantum analogue of the XOR gate. Figure \ref{cnot}a outlines why this operation is difficult:  The two optical paths that encode the target qubit are combined at a 50\% reflecting beamsplitter (BS). The output modes of this BS are then combined at a second BS to form a Mach-Zehnder interferometer. The logical operation of this interferometer by itself is to do nothing: classical interference of the single photon in the interferometer results in the target photon exiting in the same state it entered in, \emph{i.e.} $|0\rangle\rightarrow|0\rangle$; $|1\rangle\rightarrow|1\rangle$. If, however, the $\pi$ phase shift is applied inside the interferometer (such that $|0\rangle+|1\rangle\leftrightarrow|0\rangle-|1\rangle$) the target qubit undergoes a bit-flip or NOT operation: $|0\rangle\leftrightarrow|1\rangle$. A CNOT must implement this phase shift if the control photon is in the ``1" path. No known or foreseen nonlinear optical material has a non-linearity strong enough to implement this conditional phase shift (although progress has been made with single atoms in high-finesse optical cavities \cite{tu-prl-75-4710,ao-nat-443-671}, 
as discussed below, and electromagnetically induced transparency has been considered \cite{sc-oe-21-1936}).

In 2001 a surprising breakthrough showed that scalable quantum computing is possible using only single photon sources and detectors, and linear optical networks \cite{kn-nat-409-46}---\emph{i.e.} without the need for an optical nonlinearity. 
This scheme uses additional auxiliary (or `ancilla') photons that are not part of the computation, but enable a CNOT gate to work.
A cartoon of a non-deterministic (probabilistic with success signal)  CNOT is shown in Fig. \ref{cnot}b. The control and target qubits, together with two auxiliary photons, enter a (here unspecified) optical network of BSs (essentially a multi-path nested interferometer), where the four photons' paths are combined, and quantum interference can occur (Fig.~\ref{cnot}c). At the output of this network, the control and target photons emerge, having had the CNOT logic operation applied to their state, conditional on a single photon being detected at both detectors. This detection event occurs with probability $P<1$; the rest of the time another detection pattern is recorded. 
The success probability of this nondeterministic CNOT can be boosted to near-unity by harnessing quantum teleportation \cite{be-prl-70-1895,bo-nat-390-575}---a process whereby the unknown state of a qubit can be transferred to another qubit. The idea is to teleport the control and target qubits onto the output of a non-deterministic gate that has already been measured to have worked \cite{go-nat-402-390}. 

Although this `KLM' scheme \cite{kn-nat-409-46} was `in-principle' possible, initially the large resource overhead arising from the nondeterministic interactions and the difficulty of controlling photons moving at the speed of light made it practically daunting. This situation has changed over the past several years \cite{ob-sci-318-1567}: Experimental proof-of-principle demonstrations of two- \cite{ob-nat-426-264,ob-prl-93-080502,pi-pra-68-032316,ga-prl-93-020504} and three-qubit gates \cite{la-nphys-5-134} 
were followed by demonstrations of simple-error-correcting codes\cite{pi-pra-71-052332,ob-pra-71-060303,lu-pnas-08122008} and small-scale quantum algorithms\cite{lu-prl-99-250504,la-prl-99-250505}. New theoretical schemes, which dramatically reduced the considerable resource overhead\cite{yo-prl-91-037903,ni-prl-93-040503,br-prl-95-010501,ra-prl-95-100501} by applying the previously abstract ideas of cluster state (or measurement-based) quantum computing\cite{ra-prl-86-5188}, were soon followed by experimental demonstrations\cite{wa-nat-434-169,pr-nat-445-65}.

Even with these advances, the resource overhead associated with non-deterministic gates remains high. 
An alternative approach is to interact photons deterministically via an atom-cavity system \cite{pe-prl-75-3788,du-prl-92-127902,tu-prl-75-4710}, which can be configured to implement arbitrary deterministic interactions \cite{de-pra-76-052312,st-pra-78-032318} 
(semiconductor approaches are discussed below). However, there may be a payoff between resource overhead and susceptibility to errors. 
Whatever the approach to implementing gates,  the realisation of multiple high-fidelity deterministic single-photon sources remains a major challenge. In the demonstrations described above, single photons were generated via the process of spontaneous parametric downconversion (SPDC): a bright `pump' laser is shone into a non-linear crystal, aligned such that a single pump photon can spontaneously split into two ``daughter" photons, conserving momentum and energy. 
Multiplexing several (waveguide) SPDC sources \cite{mi-pra-66-053805} could provide an ideal photon source, or single atom or atom-like emitters, such as the semiconductor quantum dots described below, could be used. The latter shows potential for emitting a string of photons pre-entangled in a cluster state \cite{li-prl-103-113602}, and as nodes in quantum networks.

\vspace{12pt}
\noindent\textbf{Quantum metrology and lithography}\\
All science and technology is founded on measurement. Improvements in precision have led not just to more detailed knowledge but also new fundamental understanding. This quest to realize ever more precise measurements raises the question are there fundamental limits. Since a measurement is a physical process one may expect the laws of physics to enforce such limits. This is indeed the case, and it turns out that explicitly quantum mechanical systems are required to reach these limits \cite{gi-sci-306-1330}.

The subwavelength precision offered by optical phase~$\phi$ measurements in an interferometer (Fig. \ref{cnot}d) is the reason that they have found applications across all fields of science and technology, from cosmology (gravity wave detection) to nanotechnology (phase-contrast microscopy). 
Despite this great sensitivity, there is a limit: 
For a finite resource (energy, number of photons, \emph{etc.}) the phase sensitivity is limited by statistical uncertainty.
It has been shown that using semi-classical probes (\emph{eg.} coherent laser light) limits
the sensitivity $\Delta\phi$ to the standard quantum limit (SQL): $\Delta\phi\sim1/\sqrt{N}$, where $N$ is the average number of photons used \cite{ca-prd-23-1693,yu-pra-33-4033,gi-prl-96-010401}. The more fundamental Heisenberg limit is attainable with the use of a quantum probe (\emph{eg.} an entangled state of photons): $\Delta\phi\sim1/N$ (Refs. \onlinecite{gi-sci-306-1330,gi-prl-96-010401})---quantum metrology. 

The Heisenberg limit and the SQL can be illustrated with reference to an interferometer (Fig. \ref{cnot}d), where we represent a single photon in mode $a$ and no photons in mode $b$ by the quantum state $|10\rangle_{ab}$. After the first beam splitter this photon is in a quantum mechanical superposition of being in both paths of the interferometer: $(|10\rangle_{cd}+|01\rangle_{cd})/\sqrt{2}$. After the $\phi$ phase shift in mode $d$, this superposition evolves to the state $(|10\rangle_{cd}+e^{i\phi}|01\rangle_{cd})/\sqrt{2}$. After recombining at the second beam splitter, the probability of detecting the single photon in mode $e$ is $P_{e}=(1-\sin\phi)/2$ (this is just classical interference at the single photon level). Determination of $P_e$ can therefore be used to estimate an unknown phase shift $\phi$. If we repeat this experiment $N$ times then the uncertainty in this estimate is $\Delta\phi\sim1/\sqrt{N}$---the SQL, arising from a Poissonian statistical distribution (the same limit is obtained when a bright laser and intensity detectors are used). 

If, instead of using photons one at a time, we were able to prepare the maximally entangled $N$-photon `NOON' state $(|N0\rangle_{cd}+|0N\rangle_{cd})/\sqrt{2}$ inside the interferometer, this state would evolve to $(|N0\rangle_{cd}+e^{iN\phi}|0N\rangle_{cd})/\sqrt{2}$ after the $\phi$ phase shift. From this state we could estimate the phase with an uncertainty $\Delta\phi\sim1/N$---the Heisenberg limit---an improvement of $1/\sqrt{N}$ over the SQL. Beating the SQL is known as phase super-sensitivity 
\cite{mi-nat-429-161,re-prl-98-223601}.

Interference experiments using two- \cite{ou-pra-42-2957}, 
three- \cite{mi-nat-429-161}, and four-photon states \cite{wa-nat-429-158,na-sci-316-726} 
have been demonstrated, giving rise to a detection probability $p \propto \sin(N \phi)$. Observation of such a ``$\lambda/N$" fringe, with a period $N$ times shorter
than a conventional interferometer with wavelength $\lambda$ light, is called \textit{phase super-resolution} \cite{mi-nat-429-161}. 
However, it has been demonstrated that phase super-resolution can be observed using only semi-classical resources \cite{re-prl-98-223601}.
Therefore observation of $\lambda/N$ fringes does not guarantee quantum enhanced phase sensitivity and precise accounting of resources is required \cite{ok-njp-10-073033}. 

The closely related idea of quantum lithography involves using quantum states of light, such as the NOON state, to harness the ``reduced de-Broglie" wavelength to lithographically define $\lambda/2N$-sized features \cite{bo-prl-85-2733}. 
Significant challenges include achieving arbitrary two-dimensional patterns and realizing N-photon resists. For quantum metrology, it is important to consider whether the phase to be measured is fixed but unkown, or time varying, requiring a high bandwidth measurement. A recent breakthrough showed that the need for complicated entangled states could be replaced by increased measurement time \cite{hi-nat-450-393}, 
which is useful in the former case. Gravity wave detection is an example of the latter case, which can best be addressed by the CV approaches described below.

\vspace{12pt}
\noindent\textbf{Quantum technologies with bright laser beams}\\
The same non-linear crystal used in SPDC can be used to deterministically create quantum states of a bright laser beam: The variance in the generalized amplitude $x$ and phase $p$ of a light beam are bound by the quantum uncertainty relation: $\Delta x\Delta p\ge\hbar/2$. The output of a laser has $\Delta x=\Delta p$; while a `squeezed' state of light  has $\Delta x\neq\Delta p$.
Squeezed states include a beam of only even numbers of photons ($\sum_{n=0}^{\infty} c_n |2n \rangle$, where $n$ is the photon number);
and entangled two-mode squeezed vacuum ($\sqrt{1-q^2} \sum_{n=0}^{\infty} q^n |n \rangle_A |n \rangle_B$ where $q=\tanh r$ and $r$ is the squeezing parameter).
Such squeezed states can be used as an alternative to the (discrete, two-level) qubit encoding described above. 
As with single photons, quantum entanglement for CV photonic quantum technologies can be created in several degrees of freedom of light: the most common is amplitude and phase quadratures \cite{ou-prl-68-3663}; and others include polarization \cite{bo-prl-89-253601,ko-pra-65-052306,la-pra-71-022313} and spatial modes \cite{wa-sci-321-541}.

CV quantum communication can be regarded as a quantum version of conventional coherent communication, where information is encoded in coherent states of light---laser light. 
The essence of CV quantum communication is an `optimum measurement', which projects the encoded states onto some entangled basis states, and gives us channel capacity beyond the Shannon limit \cite{sa-pla-236-1}. The realisation of this measurement can be regarded as QIP, and so CV quantum communication and QIP are inseparable and since the processing must include coherent states of light it is CV QIP. 
Quantum metrology schemes using adaptive homodyne measurement \cite{wi-prl-75-4587} have been demonstrated \cite{ar-prl-89-133602}. This type of `quantum feedback and 
control' is becoming a powerful tool for quantum metrology.

The most fundamental component of CV photonic quantum technologies is CV quantum teleportation \cite{Vaidman94,Braunstein98,Furusawa98}. 
The fidelity $F$ of CV teleportation is directly determined by the amount of squeezing (typically quantified by the reduction in noise level of the squeezed variable below the unsqueezed shot noise value, measured in dB) of the quantum entanglement resource: $F\le(1+e^{-2r})^{-1}$.
Achieving strong squeezing is experimentally challenging because losses destroy the even-photon nature and an infinite level of squeezing is not physically possible---it would require an infinite amount of energy (number of photons).
The longstanding record of 6dB of squeezing \cite{Polzik92} was overcome using periodically poled KTiOPO$_4$ (PPKTP) as the nonlinear medium in a subthreshold optical parametric oscillator (OPO) cavity to achieve 7dB of squeezing \cite{Suzuki06}, and 9dB with improvement of phase stability in the homodyne measurement \cite{Takeno07}. In 2008 10dB was achieved with a monolithic MgO:LiNbO$_3$ OPO \cite{Vahlbruch08}, which would correspond to a teleportation fidelity of 0.91.
In actual teleportation experiments a fidelity of 0.83 has been achieved \cite{Yukawa08}, equivalent to 7dB of effective squeezing.

The advantage of QIP with single photon qubits is the near-unit fidelity of operations; however, the lack of a strong optical non-linearity at the single photon level means that one has to select success events after the processing (as described above) making processing inevitably slow.
By contrast, the advantage of QIP with CVs is the deterministic or unconditional nature of processing;  while the major disadvantage is non-unit fidelity of the processing because of the impossibility to achieve an infinite amount of squeezing ($r < \infty$).
Thus hybridization of qubits and CVs for photonic QIP could be desirable for the realization of QIP with unit fidelity and high success rate.

\begin{figure}[t]
\begin{minipage}[t]{0.02\textwidth}
            \vspace{0pt}
            \textbf{a}
        \end{minipage}
                \begin{minipage}[t]{0.45\textwidth}
            \vspace{0pt}
\includegraphics[width=5.5cm]{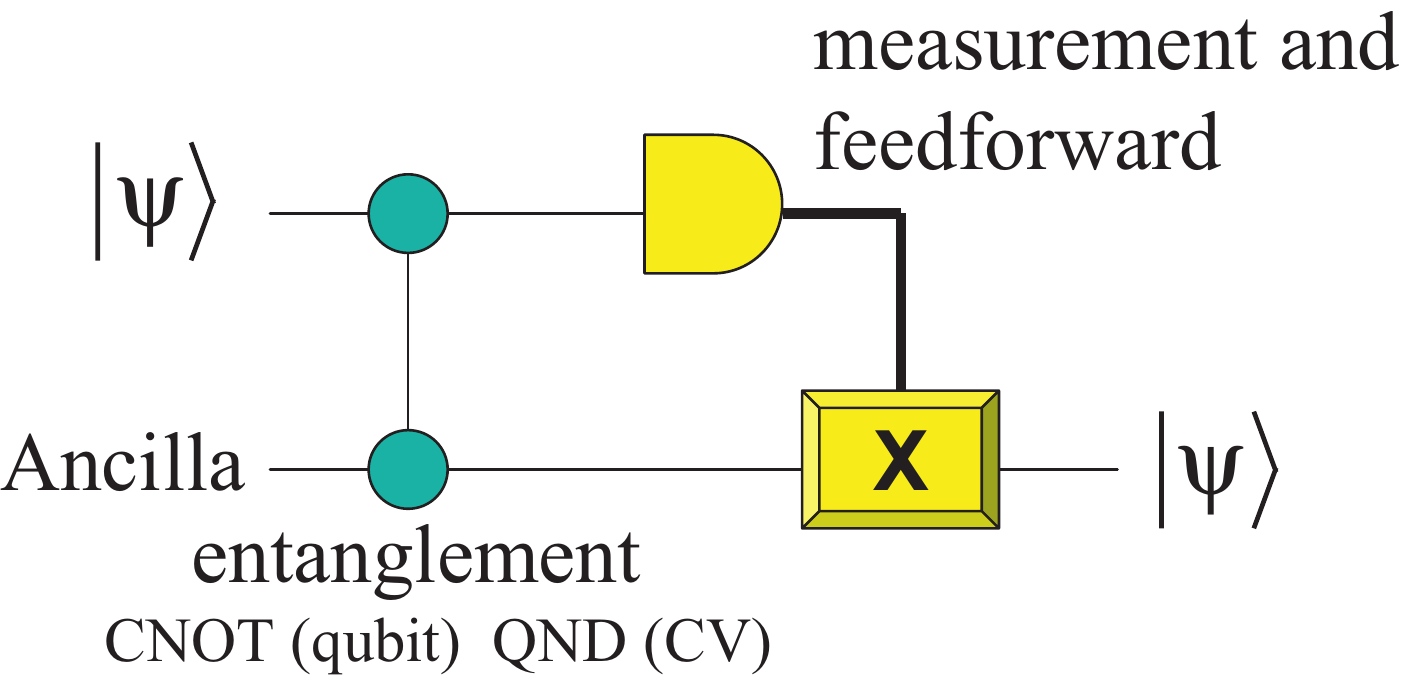}
        \end{minipage}\\

\begin{minipage}[t]{0.02\textwidth}
            \vspace{0pt}
            \textbf{b}
        \end{minipage}
                \begin{minipage}[t]{0.45\textwidth}
            \vspace{10pt}
\includegraphics[width=5.5cm]{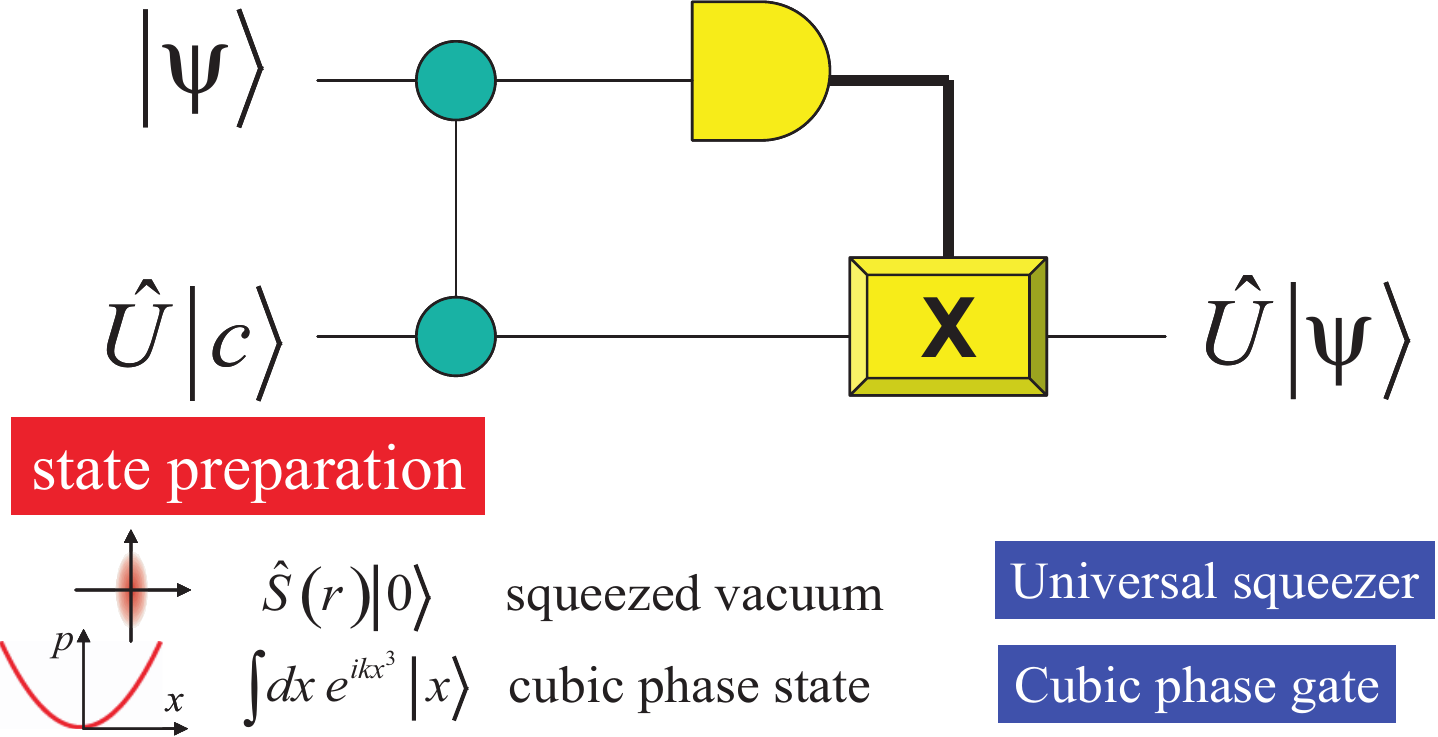}
        \end{minipage}\\
        
\begin{minipage}[t]{0.02\textwidth}
            \vspace{0pt}
            \textbf{c}
        \end{minipage}
                \begin{minipage}[t]{0.45\textwidth}
            \vspace{10pt}

\includegraphics[width=6.5cm]{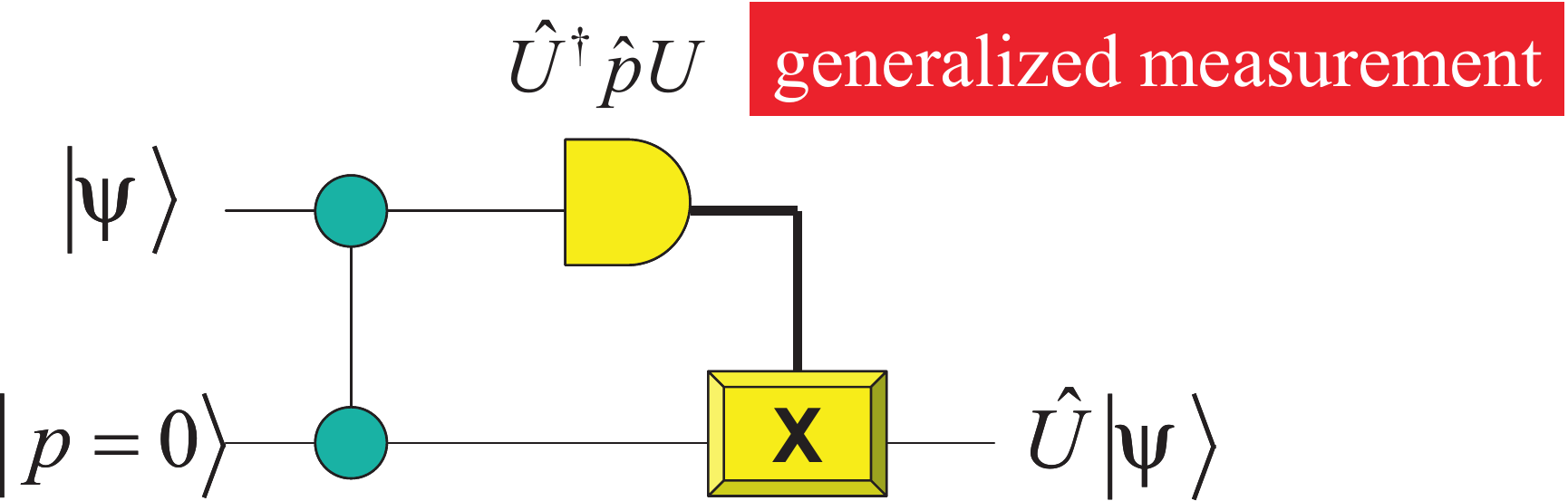}
        \end{minipage}
        
\caption{\textbf{a}, Generalized quantum teleportation. \textbf{b}, Off-line quantum information processing with the generalized quantum teleportation. \textbf{c}, Generalized quantum teleportation is applied to one-way quantum computation with cluster states. A measurement is generalized as $\hat{U}^{\dagger} \hat{p} \hat{U}$.}
\label{gen-tel}
\vspace{-0.4cm}
\end{figure}

\vspace{12pt}
\noindent\textbf{Encoding in `Schr\"odinger kittens'}\\
A squeezed vacuum created by SPDC is a beam of even number of photons; when a single photon or two photons are subtracted from it, the resulting state is a `Schr\"odinger kitten'---a superposition of coherent states of opposite phase $|\pm \alpha \rangle\equiv| \alpha \rangle \pm | \! -\alpha \rangle$, where $\alpha \sim 1$ (Ref. \onlinecite{Dakna97}). These `kittens' are almost orthogonal to each other and can be used as logical qubits: $|0 \rangle_{\rm L} = |- \alpha \rangle$; $|1 \rangle_{\rm L} = |+ \alpha \rangle$.
Since $|\pm \alpha \rangle$ are CV states, this can be regarded as hybrid qubit-CV QIP. These Schr\"odinger kittens have been created in the lab \cite{Ourjoumstev06,Neergaard06,Takahashi08}. 

Squeezing bandwidth is the most important factor for handling these `kittens'. This is because the avalanche photodiodes typically used for the photon subtraction have a much wider bandwidth than that of the squeezer, and thus bandwidth of the `kittens' is the full bandwidth of the squeezer. To handle the `kittens', the bandwidth of QIP should be broader than that of the `kittens'. More generally the bandwidth determines the speed of QIP. Since a cavity is usually used for the enhancement of nonlinearity to achieve high level of squeezing, the bandwidth is limited by the cavity bandwidth: $\sim$100 MHz at most. For broadband quantum teleportation and QIP, we should therefore not use a cavity. An alternative is to use a waveguide to enhance the nonlinearity, which has been done with waveguided periodically poled LiNbO$_3$ (PPLN) \cite{Yoshino07}. The bandwidth of squeezing and entanglement in this case is only limited by the bandwidth of the phase matching condition in principle: $\sim$10THz.

Finally, single photons can be created via single photon subtraction from weekly squeezed vacuum, or so-called `photon pairs' (two photons minus one photon gives one photon). Again, to handle single photon polarisation qubits, the bandwidth of QIP should be broader than that of the single photons. In this way one can handle polarised-photon qubits in a CV context, if the bandwidth is broad enough. This is hybridisation of qubits and CVs, the first step of which was recently demonstrated via CV teleportation of `Schr\"odinger kittens' created with photon subtraction \cite{Lee09}.

\begin{figure}[t]

\begin{minipage}[t]{0.02\textwidth}
            \vspace{0pt}
            \textbf{a}
        \end{minipage}
                \begin{minipage}[t]{0.45\textwidth}
            \vspace{0pt}

\includegraphics[width=0.95\linewidth]{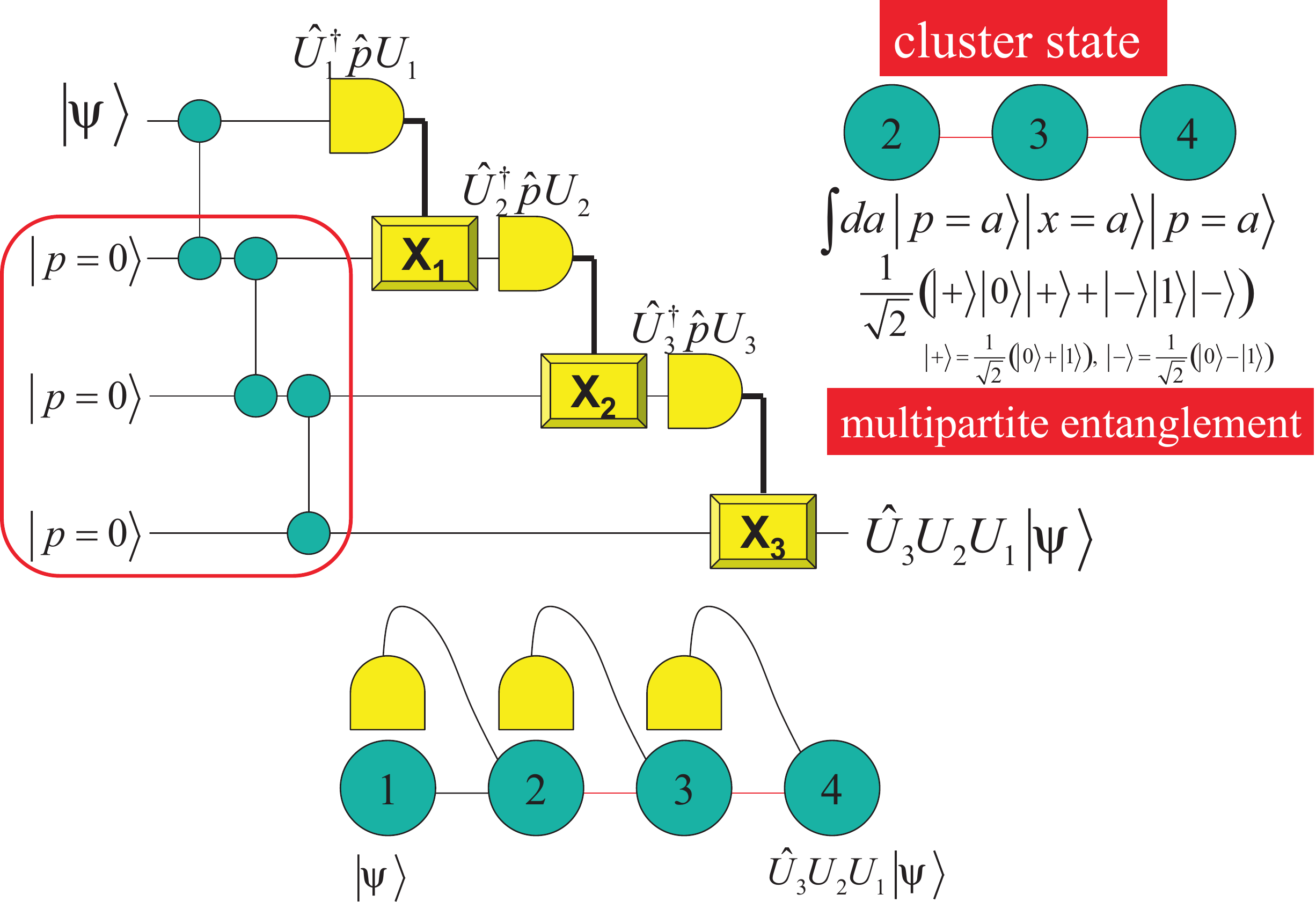}
        \end{minipage}\\

\begin{minipage}[t]{0.02\textwidth}
            \vspace{0pt}
            \textbf{b}
        \end{minipage}
                \begin{minipage}[t]{0.45\textwidth}
            \vspace{10pt}

\includegraphics[width=0.75\linewidth]{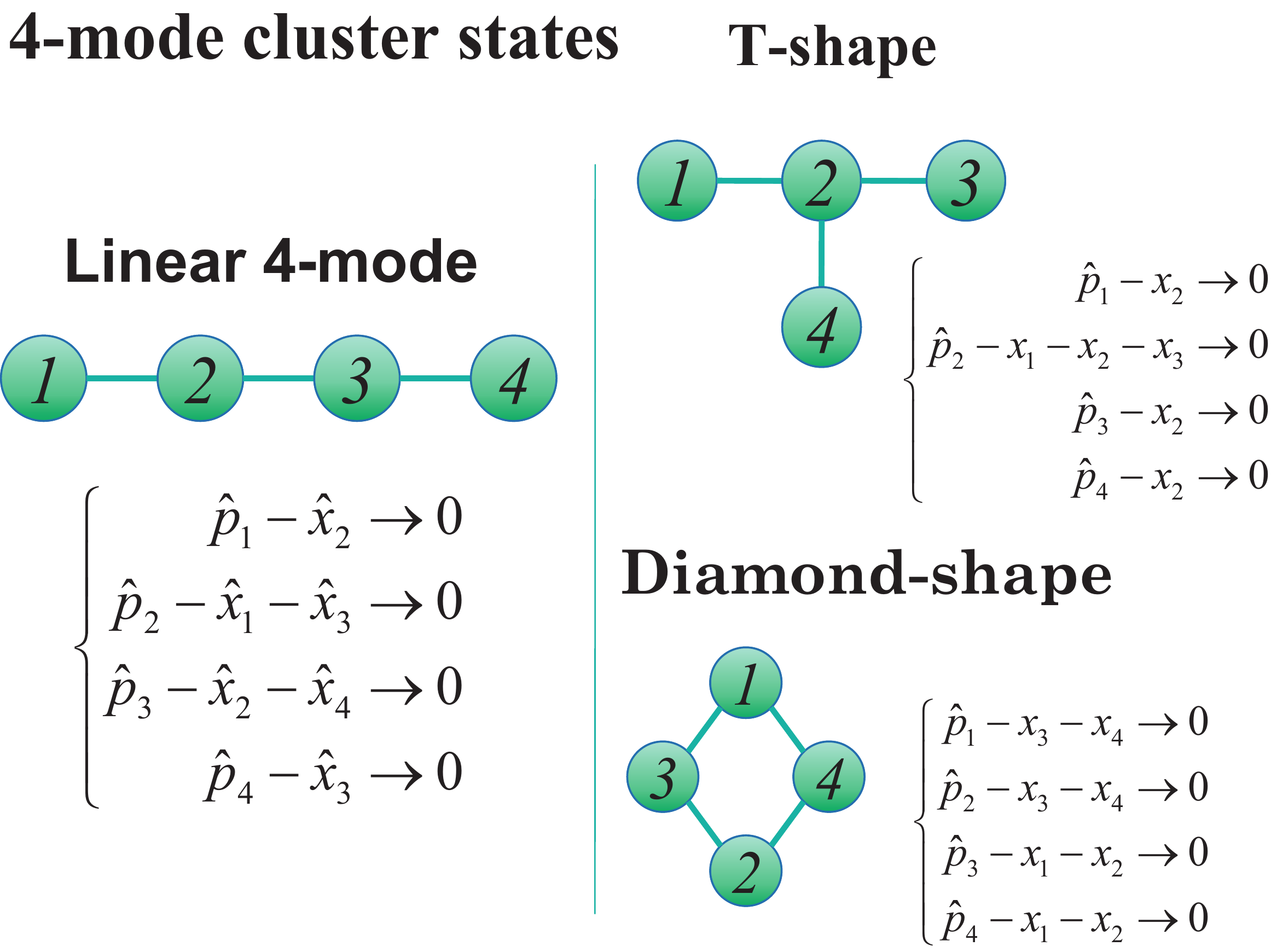}
        \end{minipage}
\caption{\textbf{a}, One-way quantum computation with cluster states. This example is on a three-mode \textit{linear} cluster state. \textbf{b}, Experimentally created CV cluster states \cite{Yukawa08c}. These are simultaneous eigenstates of the operators listed near the figures. 
For qubits, a cluster state is an eigenstates of so-called `stabilizers':
$\hat{\sigma}_x^{(i)} \bigotimes_{i^{\prime} \in N(i)} \hat{\sigma}_z^{(i^{\prime})}$,
where $\hat{\sigma}_x$ is Pauli-X operator, $N(i)$ is the set of nearest neighbor qubits of qubit $i$, and $\hat{\sigma}_z$ is Pauli-Z operator \cite{Raussendorf01}.
For CVs, they are again the eigenstates of stabilizers:
$\hat{X}_i (s_i) \prod_{i^{\prime} \in N(i)} \hat{P}_{i^{\prime}} (s_i)$,
where $\hat{X}(s)$ and $\hat{P}(s)$ are the $x$- and $p$-translation operators, respectively, and $N(i)$ is the set of nearest neighbor modes of mode $i$ \cite{vanLoock07}.
For example, the three-qubit linear cluster state is $(|~\!\!\!\!+~\!\!\!0~\!\!\!+~\!\!\!\!\rangle + |~\!\!\!\!-~\!\!\!1~\!\!\!-~\!\!\!\!\rangle)/\sqrt{2}$, where $|\pm \rangle = (|0 \rangle \pm |1 \rangle)/\sqrt{2}$,
and the three-mode linear CV cluster state is
$\int_{-\infty}^{\infty} da |p=a \rangle |x=a \rangle |p=a \rangle$.}
\label{cluster}
\vspace{-0.4cm}
\end{figure}

\vspace{12pt}
\noindent\textbf{Generalized quantum teleportation}\\
The concept of quantum teleportation has been extended to \textit{generalized} quantum teleportation\cite{Zhou00,Menicucci06}  (for both qubit and CV regimes), which can be applied to off-line QIP. The essence of the scheme (Fig.~\ref{gen-tel}a) is \textit{off-line}: pre-existing entanglement between the input $|\psi \rangle$ and an ancilla followed by measurement and feedforward.
An example is shown in Fig.~\ref{gen-tel}b in which the ancilla is a specific state $\hat{U} |c \rangle$, where the unitary operation $\hat{U}$ is the one we want to apply on the input.
The crucial advantage of the scheme is that the difficulty of operation is confined in state preparation of the ancilla, since the fidelity of teleportation itself is rather high: one does not have to make the unitary operation for an arbitrary input, instead it is enough to make it on a particular state $|c \rangle$, which is much easier than the case for an arbitrary input.
There are two important quantum gates that make use of this scheme: a universal squeezer \cite{Filip05,Yoshikawa07} and a cubic phase gate \cite{Gottesman01}, where the ancillae are a squeezed state and a cubic phase state, respectively.
The universal squeezer \cite{Yoshikawa07} and a quantum non-demolition (QND) entangling gate with the squeezers \cite{Yoshikawa08} have been demonstrated.
Here the QND entangling gate is the CV version of a CNOT gate which is also very important for CV QIP.

Another key application of generalized quantum teleportation is one-way quantum computation with cluster states both in qubit and CV regimes \cite{Menicucci06,ra-prl-86-5188} (Fig. \ref{cluster}).
The essence of the scheme is `generalized measurement' as shown in Fig.~\ref{gen-tel}c. The difference between the schemes in Figs.~\ref{gen-tel}b and c is that the ancilla  in Fig.~\ref{gen-tel}b is always $|p=0 \rangle$\  (eigenstate of $\hat{p}$ with zero eigenvalue). Measurement in Fig.~\ref{gen-tel}c is generalized as $\hat{U}^{\dagger} \hat{p} \hat{U}$ which corresponds to projection onto eigenstates of the operator $\hat{U}^{\dagger} \hat{p} \hat{U}$.

The scheme of Fig.~\ref{gen-tel}c can be cascaded as shown in Fig.~\ref{cluster}a. In this case, one can prepare some special entangled state of many parties (optical beams) called a `cluster state' in advance as shown in Fig.~\ref{cluster}b. By making a measurement on $\hat{U}_i^{\dagger} \hat{p} \hat{U}_i$ at each mode (optical beam) according to the desired operation and feeding the results forward, one can get the desired output state $\hat{U}_3 \hat{U}_2 \hat{U}_1 |\psi \rangle$. In some sense, the measurements on $\hat{U}_i^{\dagger} \hat{p} \hat{U}_i$ are the `software' and one can get desired outputs by just changing the `software'.
Toward this goal, Menicucci \textit{et al.} proposed an efficient way to generate multimode CV cluster states \cite{Menicucci08}. 
Moreover, the CV cluster states shown in Fig. \ref{cluster} have been generated experimentally \cite{Su07,Yukawa08c}.

\vspace{12pt}
\noindent\textbf{Photonics for Quantum Technologies}\\
Our ability to generate, control, and detect light has been driven by, and now permeates, all fields of human activity from communication to medicine. Generating, detecting, and manipulating quantum states of light (including single photons and squeezed states) is more challenging than for bright laser beams, but many techniques can nevertheless be adapted from the rich field of photonics. The most challenging tasks in quantum optical circuits are requirement for quantum interference of two (or more) photons and achieving interaction between two photons (as is required for nontrivial two qubit gates). The former challenge results from imperfections in the processes used for single photon generation, and the latter from the fact that optical nonlinearities are generally very small or negligible at a single photon level.

The impressive proof-of-principle demonstrations of photonic quantum technologies described above have mostly relied on large-scale (\emph{bulk}) optical elements (beamsplitters, mirrors, \emph{etc.}) bolted to room-sized optical tables, with light propagating in air. In addition, single photon qubit approaches have relied on unscalable single photon sources and detectors. For both single photon qubit, and bright CV approaches there is now a need to develop high performance sources, detectors and optical circuits, ideally integrated on a single optical chip. For single photon approaches it is also desirable to realise a strong optical non-linearity at the single photon level, while for CV approaches an integrated high-bandwidth squeezer is desirable. 

These tasks lie at the interface between quantum optics, device (nano) fabrication and photonics. In this respect, the mature field of photonics has much to offer the relatively immature field of optical quantum technologies. Already there are important examples, including photonic quantum circuits on a silicon chip \cite{po-sci-320-646}, 
high efficiency photon number resolving detectors\cite{spd}, semiconductor cavity quantum dot single photon sources \cite{ref:Charlie}, and photonic crystal quantum dot based single photon non-linearities \cite{faraon-DIT}. We now take a brief look at recent developments in these areas.

\begin{figure}[t!]
\includegraphics[width=\linewidth]{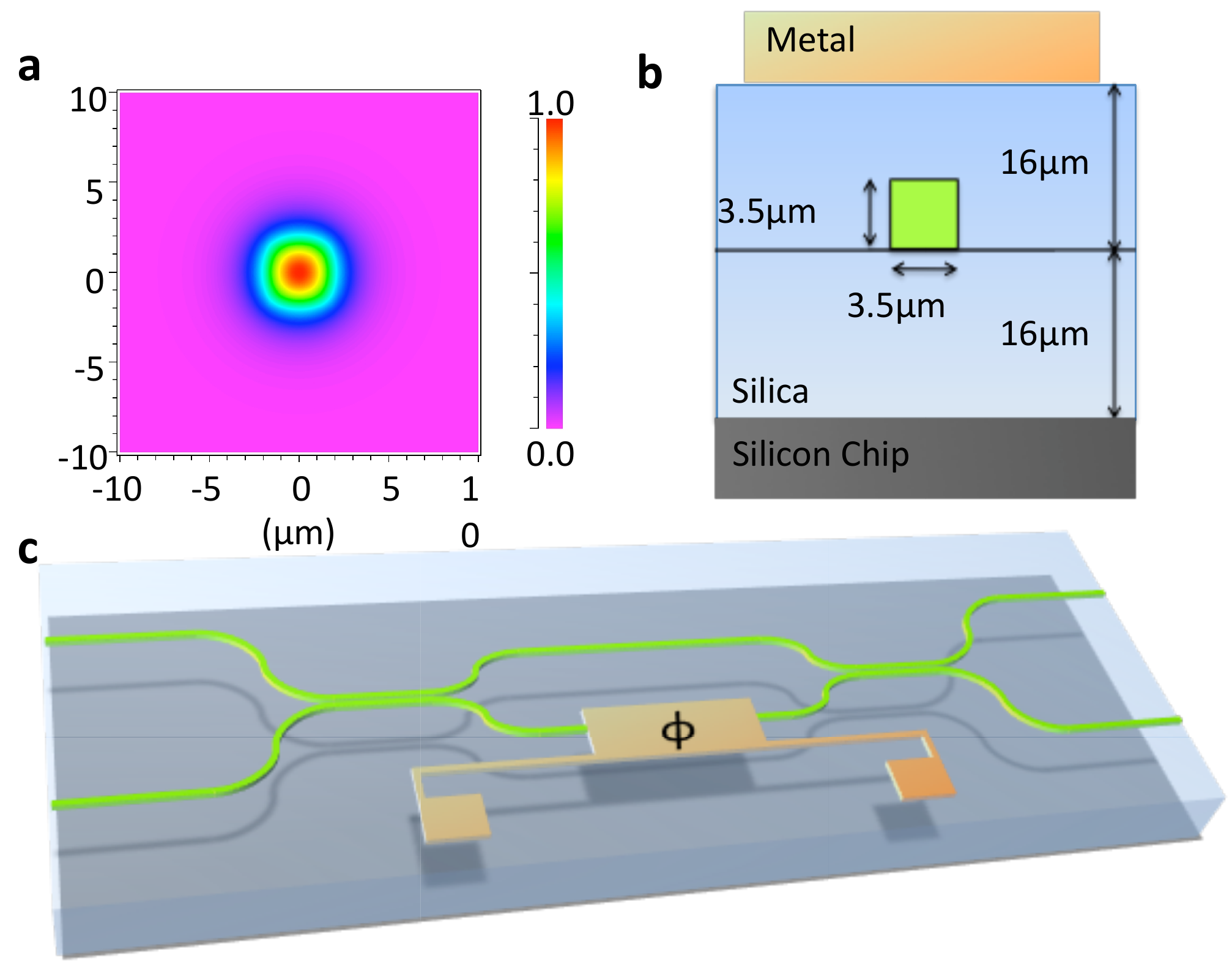}
\caption{\textbf{a}, Light is guided in a waveguide much like in an optical fibre in this simulation of the transverse intensity profile of single mode propogation. \textbf{b}, Silica-on-silicon waveguide architecture \textbf{c}, A waveguide Mach-Zehnder interferometer in which waveguide directional couplers can replace the bulk optics beamsplitters of Fig. \ref{cnot}. A metal element forms a resistive heater that locally changes the refractive index and thereby the phase in one waveguide of the interferometer.}
\label{waveguides}
\vspace{-0.5cm}
\end{figure}

\vspace{12pt}
\noindent\textbf{Integrated quantum optical circuits}\\
A promising approach to miniaturizing and scaling optical quantum circuits is to use the on-chip integrated waveguide approach developed primarily for the telecommunications industry and used for stable time-bin interferometers in QKD demonstrations at 1550 nm (Refs. \onlinecite{ho-ol-29-2797,ta-pra-72-041804}). Such an approach promises to improve performance since spatial mode matching, crucial for classical and quantum interference, should be near-perfect in such an architecture. Recently, silica-on-silicon waveguide quantum circuits were fabricated and shown to perform quantum logic gates with high fidelity (Fig.~\ref{waveguides})\cite{po-sci-320-646}. Integration of controlled phase shifters in integrated interferometers have been used to control single photon qubit states as well as manipulate multi-photon entangled states of up to four photons, including a demonstration of quantum metrology on a chip \cite{matthews-2008}. An integrated quantum optical circuit consisting of several one- and two-qubit gates was recently used to perform a compiled version of Shor's quantum factoring algorithm on a chip \cite{po-sci-325-1221}.

An alternative fabrication technique based on laser direct writing has been demonstrated \cite{marshall-2008}. It promises rapid prototyping, high-density three-dimensional devices, fabrication in material systems that may not be amenable to conventional lithography, and provides great control over the transverse spatial mode---important for low loss coupling to sources or detectors. A hybrid fabrication approach using UV direct write has also been demonstrated \cite{sm-oe-17-13516}.
Waveguide squeezers  have been used to create entangled-beams for CV systems \cite{Yoshino07}. 
A number of challenges remain to be addresses, including low-loss interfacing with sources, detectors and optical non-linearities, further miniaturisation, fast switching and reconfigurable circuits (via and electro-optic effect for example).

\vspace{12pt}
\noindent\textbf{Detectors}\\
A detailed discussion of photo-detectors for quantum technologies is beyond the scope of this review; here we outline some key points. Since photodiodes show almost unit quantum efficiency for visible and near IR wavelength, balanced homodyne detectors with photodiodes can be almost ideal quantum detectors for CV systems. However, it is known that universal quantum operation is not possible only with squeezed states of light, linear optics, and homodyne detection \cite{lo-prl-82-1784}. To get the universality of CV QIP (and also qubit QIP), we need a higher order nonlinearity which can be obtained via the measurement-induced nonlinearity described above. For the meaurement-induced nonlinearity, photon counting is essential for both CVs\cite{Gottesman01} and qubits. Moreover, there is a possibility to `synthesize' a powerful nonlinear measurement with `quantum feedback and control'. One example is adaptive homodyne measurement mentioned above. It seems clear, therefore, that qubit, CV and hybrid approaches will all require single photon detectors. Commercially available silicon avalanche photodiodes have an intrinsic quantum efficiency of $\sim 70\%$ at 800 nm and, like photomultiplier tubes, are unable to resolve the number of photons in a pulse---a key requirement in many applications. Great progress has been made in developing high efficiency number resolving detectors \cite{spd} based on superconducting nanowires, APDs and other technologies. 
The development of high-efficiency, photon-number-resolving detectors remains a key nano-photonics challenge.

\vspace{12pt}
\noindent\textbf{Semiconductor-based single photon sources}\\
Many quantum technologies, including quantum key distribution (QKD) and quantum computation and networking based on photonic qubits \cite{ref:Cirac,ref:Duan} require sources of single photons on demand. As the name says, such a device produces only one photon at its output when it is excited by some means (optically or electrically). Ideally, the source should have a high efficiency (i.e., a photon should be emitted and collected in each excitation cycle), small probability of emitting more than one photon per pulse (measured by the second-order coherence function $g^{(2)}(0)$), and should produce indistinguishable photons at its output (which can thus interfere with each other). These three parameters are critical for almost all quantum information processing applications, although some QKD protocols such as BB84 do not require indistinguishable photons.

The basic idea employed to generate single photons on demand is very simple: this can be done by performing pulsed excitation of any single quantum emitter (such as a quantum dot (QD), an atom, a molecule, a nitrogen vacancy center in diamond, or an impurity in semiconductor), followed by spectral filtering to isolate only one photon with desired properties at the output \cite{ref:NJPspec}. (Non-solid state approaches to single photon sources are beyond the scope of this review.) For example, an optical (or electrical) pulse would generate carriers (electrons and holes) inside a quantum dot; these carriers can occupy only discrete energy levels resulting from quantum confinement and Coulomb interaction in a QD. Hence, when such carriers recombine, they produce several photons of different frequencies, and spectral filtering can be used to isolate only one of them.

While multi-photon probability suppression ($g^2(0)$) is already small for a single, isolated quantum emitter excited using the described methods, single photon efficiency and indistinguishability are poor, as photons are emitted in random directions in space and dephasing mechanisms are strong. However, both efficiency and indistinguishability can be improved by embedding a quantum emitter into a cavity (with high quality factor Q and small mode volume V), where the spontaneous emission rate of the emitter can be enhanced relative to its value in bulk (or free-space) as a result of its coupling to the cavity mode (Purcell effect).
In this case, the external outcoupling efficiency increases by increasing the fraction of photons coupled to the cavity mode that are redirected towards a particular output where they can be collected. In addition, as a result of the Purcell effect, the radiative lifetime is reduced significantly below the dephasing time, leading to an increase in the indistinguishability of emitted photons (and also an increase in the possible repetition rate of the source). The improvement in this case happens as long as the radiative lifetime is well above the relaxation time of carriers from the higher order excited states to the first excited state (jitter time which is on the order of 10-30 ps in self-assembled InAs/GaAs quantum dots). By using this approach, single photons have been generated with high efficiency and indistinguishability (of up to 81\%) by optical \cite{ref:Charlie} and electrical excitation of quantum dots in micropillar cavities \cite{ref:shields}. However, with such incoherent excitation techniques the maximum indistinguishability that one could achieve is on the order of 90\% (by using Purcell effect to tune radiative lifetime between the jitter and dephasing times). Recently, the indistinguishability of 90\% has been reported for a single photon source based on resonant optical excitation of a single QD weakly coupled to a micropillar cavity \cite{Michler-Forchel}: in this case, the jitter time limitation is overcome (as carrier relaxation from higher order states is bypassed), but dephasing still affects the source performance.

Still, achieving a perfect indistinguishability necessary for quantum computing remains to be a challenge.  To overcome this limit, one could employ cavity quantum electrodynamics (QED) and resonant excitation of the strongly coupled quantum dot - cavity system \cite{ref:kiraz}. The field of solid-state cavity QED has witnessed an exponential growth in recent years, as described below, and it is almost certain that we will see solid-state single photon sources with perfect indistinguishability in the near future. 
(Non-solid state approaches \cite{ref:NJPspec} to single photon sources are beyond the scope of this review.)

\vspace{12pt}
\noindent\textbf{Strong single-photon non-linearities on-chip}\\
As mentioned above, one of the greatest challenges in photonic quantum information processing is achieving nonlinear interaction between two photons, which is needed for nontrivial two qubit quantum gates 
and quantum nondemolition measurements of photon numbers \cite{ref:Nogue}. This is a result of the fact that optical nonlinearities are very small at a single photon level. In the past, the largest nonlinearities have been realized with single atoms strongly coupled to  resonators \cite{Harochephasegate, Turchette} and atomic ensembles \cite{ref:seharris}. However, the field of solid state cavity QED has recently witnessed a rapid progress, including the demonstrations of strong coupling regime in photoluminescence \cite{Yoshie04, SCImamogluNature,forchelyamamoto} and coherent probing of the strongly coupled quantum dot-cavity system \cite{jvnature,painternature} It has recently been shown that the same magnitude of nonlinearity can be achieved in an on-chip configuration with a strongly coupled quantum dot-nanocavity system \cite{jvscience}. Namely, inside a photonic crystal nanocavity containing a strongly coupled quantum dot, one can presently achieve a controlled phase (up to $\pi/4$) and amplitude (up to 50\%) modulation between two modes of light at the single-photon level. Finally, photon induced tunnelling and blockade have also been demonstrated in a solid state system \cite{jvnatphys}, which makes the solid state cavity QED systems comparable to their atomic physics counterparts in terms of the achievable strength of interaction \cite{kimble-blockade}.

\begin{figure}[t!]
\includegraphics[width=\linewidth]{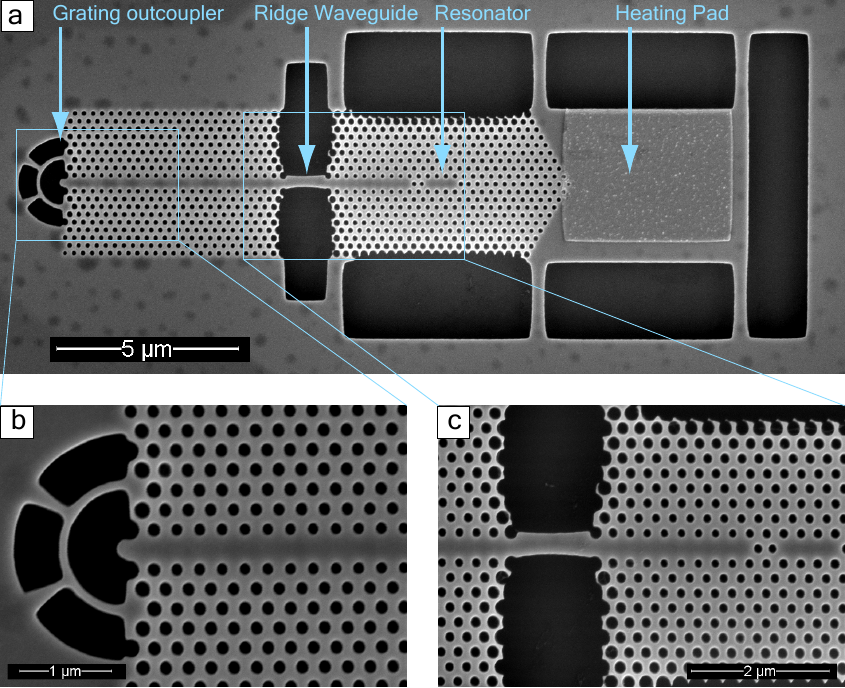}
\caption{\textbf{a}, A rudimentary photonic crystal quantum circuit (from Reference \cite{faraon-DIT}). The device
consists of a PC cavity coupled to a PC waveguide terminated with a grating outcoupler.
The cavity contains a single quantum dot strongly coupled to it. For local temperature control, the cavity is placed next to a metal pad that can be heated using an external laser beam. To increase the thermal insulation of the structure, the PC
waveguide is interrupted and a narrow ridge waveguide link is inserted. \textbf{b}, Magnified view
of the grating outcoupler. \textbf{c}, Magnified view of the ridge waveguide link}
\label{fig-PC-circuit}
\end{figure}

Although solid state cavity QED systems in many ways offer many advantages over atomic cavity QED systems (in terms of scalability, on chip architecture, miniaturization, higher speeds resulting from smaller mode volumes, and the elimination of need to trap quantum emitters), inhomogeneous broadening of solid state emitters and handling at cryogenic temperatures still pose a challenge. Several approaches to overcome these problems have been proposed, including alignment techniques for photonic crystal resonators to randomly distributed self assembled QDs \cite{badolatoimamoglu}, tuning of cavities on the whole chip by digital etching \cite{imamoglu-digitaletch} or gas condensation \cite{scherer-condensation}, local tuning of cavities by photorefractives \cite{faraon-chalcogenides}, and local tuning of QDs by temperature  \cite{faraon-QDtuning, andrei-el-tuning1} or electric field \cite{finley-QD-electrical-tuning,andrei-el-tuning2}. Many groups are also working on nitrogen vacancy in diamond for obtaining room temperature operation \cite{lukin-NVcenter, awschalom-NVcenter}, but their coupling to photonic structures is challenging, and strong coupling regime has yet to be achieved.

Researchers are also looking into quantum emitters that are compatible with telecom wavelength operation, but many of them have inferior properties relative to the emitters (QDs or NV centers) operating at shorter wavelengths. For this reason, frequency conversion techniques at a single photon level have been proposed and developed in recent years \cite{kwiat-freqconv}, 
including on chip demonstration in periodically polled lithium niobate (PPLN) waveguide geometry \cite{fejer-freqconv}

On the other hand, atomic systems are also moving towards chip scale realizations based on, e.g., silica microtoroid geometries \cite{ao-nat-443-671}. Photonic approaches not only allow for a more compact realization of quantum information processing proposals, but also enable much smaller cavity mode volumes, and higher coupling strengths between the emitters and the cavity field, thus leading to much stronger coupling regimes (and thus higher operating speeds) than previously achievable with larger scale resonators.

\vspace{12pt}
\noindent\textbf{Future Outlook}\\
We have just witnessed the birth of the first quantum technology based on encoding information in light for quantum key distribution. Light seems destined to continue to have a central role in future technologies including secure networks and quantum information processing. To date qubit and CV QIP have largely been investigated separately with much progress in each. Despite this progress, there remain many hurdles to overcome before the ultimate goal of universal QIP can be achieved. Combining these approaches may allow us to take advantage of both regimes to overcome these hurdles, particularly with respect to the power of off-line schemes based on quantum teleportation.

As we have seen, approaches to optical quantum technologies are beginning to adopt state-of-the-art developments from the field of photonics. In the near future we will likely see the development of photonic quantum technologies driving the development of photonics itself. Many challenges also remain in solid-state photonic quantum technologies. As mentioned above, indistinguishable single photons on demand have yet to be demonstrated, but as a result of the recent breakthroughs in solid-state cavity QED, we can expect developments in that direction in the near future. Moreover, although controlled phase shifts have been demonstrated between two optical beams at a single photon level, in order to reach a full $\pi$ phase shift, we need to enhance the cavity QED effects and likely also cascade several of the demonstrated elements. Finally, for the employment of these blocks in functional quantum computers and repeaters, we also need local quantum memory nodes, so combination of the demonstrated efficient photonic blocks with techniques for manipulation of solid state qubits (e.g., quantum dot or nitrogen vacancy spin states) is critical.

\noindent\textbf{Acknowledgements:} JLOB acknowledges support from EPSRC, QIP IRC, IARPA, ERC the Leverhulme Trust, and a Royal Society Wolfson Merit Award. AF acknowledges financial supports from SCF, GIA, G-COE, and PFN commissioned by the MEXT of Japan and Research Foundation for Opto-Science and Technology. JV acknowledges support from ONR, ARO, and NSF.

\end{document}